\documentclass[aps,floats,twocolumn,showpacs]{revtex4-1}
\usepackage{amssymb}
\usepackage{amsmath}
\usepackage{graphicx}

 \def\bc{\begin{center}}          \def\ec{\end{center}}
 
 \def\E{{\mathbf E}}
 \allowdisplaybreaks
\begin{document}
 \title{Natural noise and external wake field seeding in a proton-driven plasma accelerator}
 \author{K.V.Lotov$^{1,4}$, G.Z.Lotova$^{2,4}$, V.I.Lotov$^{3,4}$,  A.Upadhyay$^{5}$, T.T\"uckmantel$^{5}$, A.Pukhov$^{5}$, A.Caldwell$^{6}$}
 \affiliation{$^{1}$Budker Institute of Nuclear Physics SB RAS, 630090, Novosibirsk, Russia}
 \affiliation{$^{2}$Institute of Computational Mathematics and Mathematical Geophysics SB RAS,
630090, Novosibirsk, Russia}
 \affiliation{$^{3}$Sobolev Institute of Mathematics SB RAS, 630090, Novosibirsk, Russia}
 \affiliation{$^{4}$Novosibirsk State University, 630090, Novosibirsk, Russia}
 \affiliation{$^{5}$Institut f\"ur Theoretische Physik I, Heinrich-Heine-Universit\"at D\"usseldorf,
40225 Germany}
 \affiliation{$^{6}$Max-Planck-Institut f\"ur Physik, 80805, M\"unchen, Germany}
 \date{\today}
 \begin{abstract}
We discuss the level of natural shot noise in a proton bunch-driven plasma accelerator. The required seeding for the plasma wake field must be larger than the cumulative shot noise. This is the necessary condition for the axial symmetry of the generated wake and the acceleration quality. We develop an analytical theory of the noise field and compare it with multi-dimensional simulations. It appears that the natural noise wake field generated in plasma by the available at CERN super-protons-synchrotron (SPS) bunches is very low, at the level of a few 10 kV/m. This fortunate fact eases the requirements on the seed. Our three dimensional simulations show that even a few tens MeV electron bunch precursor of a very moderate intensity is sufficient to seed the proton bunch self-modulation in plasma.
 \end{abstract}
 \pacs{41.75.Lx, 52.35.Qz, 52.40.Mj}
 \maketitle

Particle beam-driven plasma wakefield acceleration (PWFA) is capable of producing accelerating gradients far in excess of those in conventional accelerators \cite{Nat.445-741}, but needs the drive beam to be properly shaped or compressed (see, e.g., ref.~\cite{IEEE96} for PWFA basics). So far PWFA has been experimentally studied with electron or positron beams shaped before the plasma \cite{KIPT, ANL1, Jap, Barov, PAC01-3975, PRL91-014802, Muggli, SLAC, Nat.445-741}. Recently a new approach was proposed \cite{EPAC98-806, PRL104-255003, PPCF53-014003} which assumes beam shaping by the plasma itself as a result of the transverse two-stream beam-plasma instability \cite{PoP2-1326, PoP4-1154, NIMA-410-461, PRL107-145003}. At the nonlinear stage, the instability splits the initially long beam into short bunches spaced exactly one plasma wavelength apart \cite{PoP18-024501, PoP18-103101}. Harnessing the instability would make it possible to excite strong wakefields by initially long beams without a complicated and expensive compressor or chopper. The controlled instability is the key physical effect to be demonstrated by the discussed proton-driven PWFA experiment in CERN \cite{PoP18-103101, PAC11-TUOBN5} and auxiliary experiments \cite{FACET, AIP1229-467}.

To be useful for acceleration of a witness beam, the generated wake must be axisymmetric. Yet, the plasma supports various modes of the instability, including non-axisymmetric (or hosing) ones. The latter are undesirable since they quickly destroy the beam to the state at which no strong wakefield is excited \cite{PRL104-255003}. Fortunately, if the axisymmetric mode has grown up to sufficiently high amplitude, it prevents development of other modes \cite{EPAC98-806}. Simulations show that the proper mode may dominate when an externally introduced seed perturbation is introduced \cite{PRL104-255003}. The amplitude of the perturbation must be much higher than the noise level from which hosing modes grow up. To simulate the instability correctly and to determine the required amplitude of the seed perturbation, we have to know the noise level.

There could be various sources of uncontrollable seed perturbations for instabilities. The one we are interested in is the shot noise of individual beam particles. The wakefield pattern in this case moves with the beam and will be amplified by the beam instability.

The shot noise field is a sum of wakefields left behind by separate beam particles. The contribution of a single proton located at the radius $r_b$ into the on-axis wakefield at the distance $z_b$ downstream can be taken from Ref.~\cite{PAc22-81}:
\begin{equation}\label{e1}
  E_{bz} = -2 e k_p^2 K_0(k_p r_b) \cos (k_p z_b),
\end{equation}
\begin{equation}\label{e2}
  |E_{b\perp}| = 2 e k_p^2 K_1(k_p r_b) \sin (k_p z_b),
\end{equation}
where $k_p = \omega_p/c$ is the plasma wavenumber determined by the plasma
frequency $\omega_p$ and the light velocity $c$, $e>0$ is the elementary
charge, $K_0$ and $K_1$ are the modified Bessel functions of the second kind. We use the cylindrical coordinates with the $z$-axis being the direction of beam propagation. By writing the transverse field in the form (\ref{e2}) we take into account that the wave amplitude is low and thus there is no magnetic field left behind the particle.

We will illustrate the obtained formulae by beam and plasma parameters discussed in Ref.~\cite{PoP18-103101} as SPS-LHC variant: number of beam particles $N=1.15 \times 10^{11}$, plasma density $n_p=7 \times 10^{14}\text{cm}^{-3}$, radius $\sigma_r = k_p^{-1} = 0.2\,\text{mm}$, length $\sigma_z = 12$\,mm.

\begin{figure}[t]
 \bc\includegraphics[width=234bp]{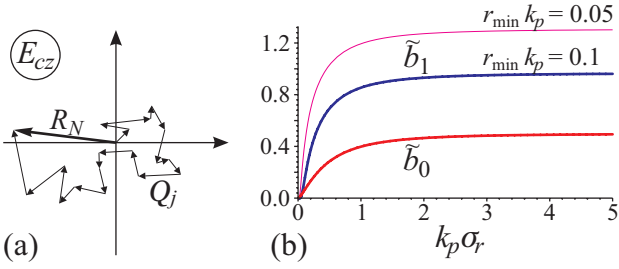} \ec
 \vspace*{-5mm}
\caption{(a) Excitation of the noise field as a random walk; (b) second moments
of probability distributions.}\label{fig1-walk}
\end{figure}
 First we calculate the longitudinal noise field $E_{nz}$ on the axis. As the beam evolves slowly as compared to the plasma timescale, the field depends on longitudinal coordinate and time in the combination $\xi = z - ct$. We consider the field as the real part of some complex function:
\begin{equation}\label{e3}
  E_{nz} (\xi) = {\rm Re}\, \left( E_{cz}(\xi) e^{i k_p \xi} \right).
\end{equation}
Each beam particle makes a contribution $Q_j$ into the complex amplitude $E_{cz}$
which depends on particle location with respect to the observation point. The
absolute value of the contribution depends on $r_b$, while the argument is determined
by $z_b$. The process of noise field excitation thus can be considered as a two-dimensional
random walk $\{R_N\}$ on the complex amplitude plane (Fig.~\ref{fig1-walk}a): $R_N=Q_1+\ldots+Q_N$.

Properties of the random walk are determined by probability
distributions of absolute value and direction of the steps. Since
the beams of interest are much longer than the plasma wavelength $2
\pi k_p^{-1}$, we can assume the isotropic distribution of the
steps. It is this assumption that rules out the collective wakefield excitation and retains only the shot noise.
The distribution of absolute values is characterized by the
probability density $f(s)$ that, in turn, is determined by the
radial distribution of beam particles:
\begin{equation}\label{e4}
  f(s) = \frac{r}{\sigma_r^2} e^{-r^2/2\sigma_r^2} \, \frac{dr}{ds},
\end{equation}
where $r(s)$ is the function reciprocal to
\begin{equation}\label{e5a}
  s(r) = 2 e k_p^2 K_0(k_p r).
\end{equation}
Thus we have to determine the expectation $E_{az} = \E |R_N|$ for large $N$ (the boldface $\E$ denotes the expectation).

Suppose that $\{B_N\}$ is a sequence of positive constants, and the distribution of $R_N/B_N$ converges weakly to the two-dimensional normal distribution with the density function
\begin{equation}\label{e6a}
 d(\nu_1,\nu_2)=\frac 1 {2\pi}\exp\bigg\{-\frac 1 2 (\nu_1^2+\nu_2^2)\bigg\}.
\end{equation}
Let $(\zeta_1,\zeta_2)$ be a random vector with the density function $d (\nu_1,\nu_2)$. Then its components are independent and have standard
normal distribution. Denote $\rho^2=\zeta_1^2+\zeta_2^2$. A straightforward calculation shows that $\rho^2$ has an exponential distribution with the density function
\begin{equation}\label{e7a}
 e(\nu)=\frac 1 2 \exp\bigg\{-\frac{\nu}{2}\bigg\},\ \ \ \ \nu>0.
\end{equation}
We have, therefore, for large $N$
\begin{multline}\label{e8a}
\E \bigg\{\bigg|\frac{R_N}{B_N}\bigg|\bigg\}\approx\E
\rho=\E\sqrt{\zeta_1^2+\zeta_2^2}= \\
= \int_0^{\infty}\sqrt{\nu}\,e(\nu)d\nu=\sqrt{\frac {\pi}{2}}.
\end{multline}
Thus, the problem reduces to finding a sequence $B_N$ that provides
the convergence to the two-dimensional normal distribution.

Let us choose an arbitrary straight line containing the origin and denote by $\varphi_j$ the angle between this direction and $Q_j$. We suppose that random variables $\varphi_j,\ j=1,2, \ldots$ have uniform distribution on $[0, 2\pi]$. Denote
\begin{equation}\label{e9a}
s_j=|Q_j|,\quad S_N=s_1\cos{\varphi_1}+\ldots+s_N\cos{\varphi_N}.
\end{equation}
We come to the one-dimensional random walk $\{S_N\}$ which is the
projection of the initial random walk $\{R_N\}$ to the chosen
direction. It is known that weak convergence of the distribution of
$S_N/B_N$ to the one-dimensional standard normal distribution for
each chosen direction guarantees us the weak convergence of the
distribution of $R_N/B_N$ to the two-dimensional standard normal
distribution (the theorem of Cram\'er -- Wold \cite{CW}). Thus, we
have the problem of one dimensional convergence. The distribution of
$s_j\cos{\varphi_j}$ is symmetric. If
\begin{equation}\label{e10a}
b_0 = \E s_1^2=\int_0^{\infty}s^2\,f(s)ds<\infty,
\end{equation}
then the distribution of $S_N/(\sigma\sqrt{N})$ is approximately
standard normal due to the classical central limit theorem, where
\begin{equation}\label{e11a}
\sigma^2=\E s_1^2\cos^2{\varphi_1}=b_0/2.
\end{equation}
Thus, the expectation of $|E_{cz}|$ after passage of $N$ particles is
\begin{equation}\label{e5}
  E_{az} =\sigma\sqrt{\frac{\pi N}{2}} = \frac{e k_p}{\sigma_r}
  \sqrt{\pi N \tilde b_0 (k_p \sigma_r)},
\end{equation}
where
\begin{equation}\label{e6}
  \tilde b_0 (k_p \sigma_r) = \int_0^{\infty}x K_0^2(x) e^{-x^2/(2 k_p^2 \sigma_r^2)}\, dx.
\end{equation}
The function $\tilde b_0(k_p \sigma_r)$ is shown in Fig.~\ref{fig1-walk}b. For the discussed parameters, $\tilde b_0 \approx 0.4$, and the expectation of the on-axis longitudinal electric field is $130$\,V/cm.

\begin{figure}[htb]
 \bc\includegraphics[width=234bp]{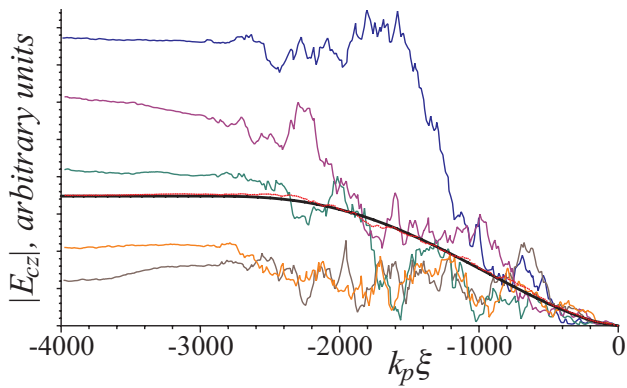} \ec
 \vspace*{-5mm}
\caption{Simulated field amplitudes for various beam ensembles (thin colored lines), the expectation value (black thick smooth line), and the field amplitude averaged over 60 beam ensembles (red dotted line) for LCODE simulations of SPS beam.}\label{fig2-ens}
\end{figure}
Figure~\ref{fig2-ens} illustrates how the actual field differs from the expectation value. These are LCODE \cite{PoP5-785} simulations of the plasma response to the full (not half-cut) SPS beam \cite{PoP18-103101}. Five thin curves correspond to five different initializations of the random number generator. The thick curve is the expectation (\ref{e5}). The dotted curve is the average over 60 different initializations of the random number generator.

\begin{figure}[htb]
 \bc\includegraphics[width=171bp]{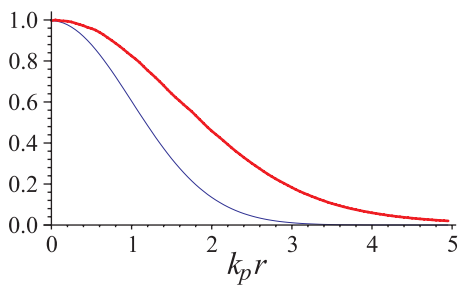} \ec
 \vspace*{-5mm}
\caption{The expected value of $|E_z(r)|$ as compared to the
expectation of $|E_z(0)|$ for $\sigma_r = k_p^{-1}$ (thick line) and
the corresponding radial profile of the beam density (thin
line).}\label{fig3-offax}
\end{figure}
The expectation for the off-axis longitudinal electric field can be found in a similar way, but with a more complicated probability density distribution used instead of (\ref{e4}). Alternatively, the expectation can be obtained by Monte Carlo simulation of the random walk at a reduced number of steps (Fig.~\ref{fig3-offax}). Note that the field area is wider than the beam itself.

\begin{figure}[htb]
 \bc\includegraphics[width=235bp]{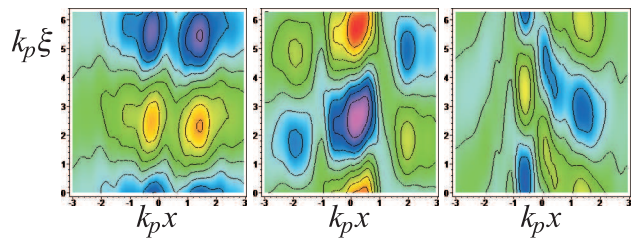} \ec
 \vspace*{-5mm}
\caption{Maps of the wakefield potential for three different beam ensembles.}\label{fig4-phi}
\end{figure}

Since the electric field excited in the plasma is a potential one, we can find
the transverse field component from the longitudinal one through the potential
$\Phi$:
\begin{equation}\label{e7}
  E_z = -\frac{\partial \Phi}{\partial z} = - i k_p \Phi, \quad
  \vec E_\perp = -\frac{\partial \Phi}{\partial \vec r_\perp}
    = \frac{1}{i k_p} \frac{\partial E_z}{\partial \vec r_\perp}.
\end{equation}
Typical portraits of the potential $\Phi$ at the plane $(x,\xi)$ are shown in Fig.~\ref{fig4-phi}. These graphs are obtained from the random walk model by summing up contributions of separate test macro-particles for three different beam ensembles. We see that the transverse scale of the potential change is shorter than the longitudinal scale. Consequently, typical transverse fields must be higher than the longitudinal ones given by (\ref{e5}).

Calculation of expectation values for transverse fields is tricky because of a diverging integral. The model of random walk with isotropically distributed steps is still applicable, but the absolute value of the step $s$ now depends of two parameters. For the field component $E_x$ we have
\begin{equation}\label{e8}
    s (r, \theta) = 2 e k_p^2 K_1(k_p r) |\cos \theta|,
\end{equation}
where $\theta$ is the polar angle of particle projection onto the $(x,y)$ plane. To find the second probability moment $b_1$ of $s$, we need the double integration:
\begin{multline}\label{e9}
    b_1 = \int s^2\,f(s)ds = \int_{r_\text{min}}^\infty dr \int_0^{2\pi} d \theta  \frac{s^2 r}{2 \pi \sigma_r^2} \, e^{-r^2/2\sigma_r^2} = \\
    =\frac{2 e^2 k_p^2}{\sigma_r^2} \int_{k_p r_\text{min}}^\infty x K_1^2 (x) e^{-x^2/(2 k_p^2 \sigma_r^2)} dx \equiv \frac{4 e^2 k_p^2 \tilde b_1}{\sigma_r^2}.
\end{multline}

\noindent  Here we limit the integration interval by some minimum radius $r_\text{min}$ to avoid divergence. At smaller radii, formula (\ref{e2}) is no longer valid. Several factors could limit the applicability of the linear cold fluid model used in derivation of (\ref{e2}).

At scales below $r_1 = n_p^{-1/3}$, the plasma cannot be considered as a continuous medium. For $n_p = 7 \times 10^{14}\text{cm}^{-3}$, $r_1 \approx 10^{-5}$\,cm.

The linear theory \cite{PAc22-81} gives the following perturbation of the plasma electron density by a single relativistic proton:
\begin{equation}\label{e10}
    \delta n_p = k_p \delta(\vec r_\perp - \vec r_b) \sin k_p z_b,
\end{equation}
where $\delta(\vec r_\perp - \vec r_b)$ is the two-dimensional delta-function. If we distribute the same amount of excess charge over the area $r_2^2$ to fulfil the linearity condition $\delta n_p \leq n_p$, then we find the scale of linearity violation by a point charge: $r_2 = \sqrt{k_p/n_p} \approx 3 \times 10^{-7}\text{cm} \ll r_1$. Therefore applicability of the linear theory is not a limitation in our case.

Nonzero temperature $T_e \sim 5$\,eV of plasma electrons modifies the plasma response on the scale of Debye length
\begin{equation}\label{e11}
    r_d \approx 743 \sqrt{\frac{T_e \, (\text{eV})}{n_p\, (\text{cm}^{-3})}} \sim 6 \times 10^{-5}\text{cm}.
\end{equation}
The modification does not take effect immediately behind the beam proton, but after thermally moving electrons have time to shift by the distance $\sim r_d$. This occurs at the time scale $\omega_p^{-1}$, so the Debye length is a limiting scale for long proton beams.

\begin{figure*}[bt]
 \bc\includegraphics[width=450bp]{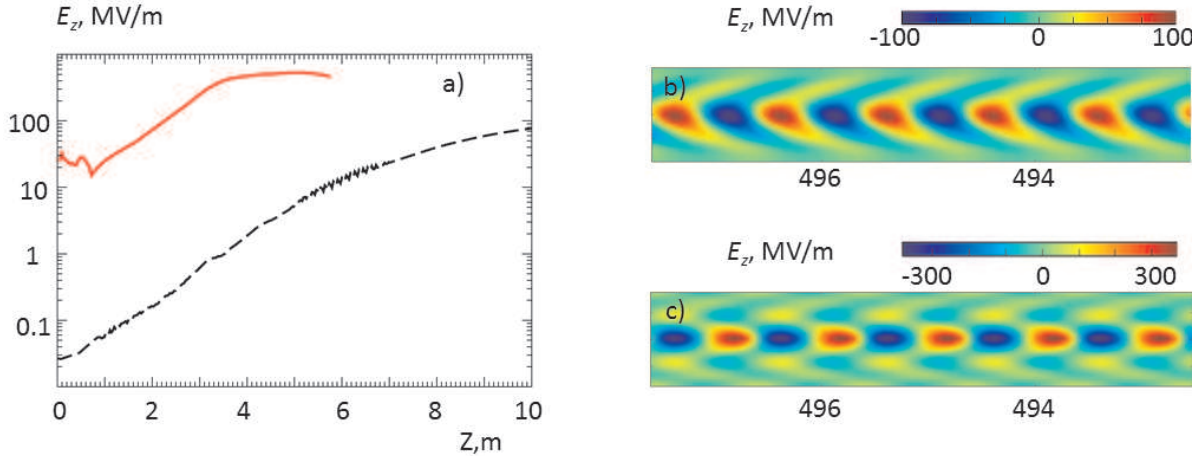} \ec
 \vspace*{-5mm}
\caption{a). Growth of the accelerationg field with propagation distance in plasmas. The broken line corresponds to an unseeded SPS proton bunch; the solid line shows the self-modulation growth when a 10 MeV electron beam has been used as a precursor, b) 2D on-axis cut of the wake field in the unseeded simulation behind the driver at the distance $z=9$~m. The field is asymmetric due to the competition between hosing and self-modulation;  c).2D on-axis cut of the wake field in the seeded simulation behind the driver at the distance $z=4$~m. The wake is symmetric.}\label{fig5}
\end{figure*}
The nonzero plasma temperature results in the nonzero group velocity of plasma waves which depends on the wavenumber $k$ of the perturbation:
\begin{equation}\label{e12}
    v_g = 3 \omega_p r_d^2 k.
\end{equation}
The sharply localized field spike (\ref{e2}) behind the proton is produced by short-wavelength wave harmonics, with transverse wavenumbers $k \sim r_3^{-1}$ being responsible for the field spike in the area of the transverse size $r_3$. The higher $k$ the faster the energy drifts out of the field spike. Assuming the time of field evolution $\sigma_z/c$, we find the minimum scale $r_3$ for which the wave has no time to drift out of the field spike:
\begin{equation}\label{e13}
    v_g \sigma_z/c = r_3, \quad r_3=\sqrt{3 k_p r_d^2 \sigma_z} \approx 3 \times 10^{-3}\text{cm} \approx 0.1\,k_p^{-1}.
\end{equation}
This is the scale we take for $r_\text{min}$. The function $\tilde b_1 (k_p \sigma_r)$ for $k_p r_\text{min} = 0.1$ is shown in Fig.~\ref{fig1-walk}b. For $\sigma_r = k_p^{-1}$, $\tilde b_1 \approx
0.85$, and the expectation of the on-axis transverse electric field component is
\begin{equation}\label{e14}
    E_{ax} = = \frac{e k_p}{\sigma_r} \sqrt{\pi N \tilde b_1 (k_p \sigma_r)} \approx 200 \, \text{V/cm}.
\end{equation}
An uncertainty in determination of $r_\text{min}$ has a little effect on the field expectation because of the weak (logarithmic) dependence of $\tilde b_1$ on $r_\text{min}$. Factor of two smaller $r_\text{min}$ results in $0.5 \ln 2$ (or 40\%) increase of $\tilde b_1$ (Fig.~\ref{fig1-walk}b) and 20\% increase of $E_{ax}$.

We have simulated the wake field by the SPS beam using the hybrid 3D particle-in-cell code VLPL \cite{JPP,hybrid}. We simulated both the "unseeded" case, when the wake field grows from the shot noise of the SPS bunch, and the case when the proton bunch self-modulation is seeded by a precursor electron bunch. The major difficulty is to simulate the shot noise properly. The number of numerical macroparticles is several orders of magnitude smaller than the $1.15\cdot 10^{11}$ protons in the SPS bunch. Thus, the noise generated by randomly seeded macroparticles would be significantly higher than the natural expectation \eqref{e5}. For this reason, we initialize all the numerical beam macroparticles at the centra of the grid cells and distribute their transverse momenta regularly and symmetric according to the bunch divergence. The longitudinal momenta have a random Gaussian distribution centered at 450~GeV/c and the spread corresponding to the given longitudinal emittance. A bunch initialized in this way generates an extremely low numerical noise field, a couple orders of magnitude below the expectation \eqref{e5}.

To adjust the noise to the natural level  \eqref{e5}, we introduce random displacement $\epsilon$  to the regular positions of macroparticles in the configuration space:

\begin{equation}\label{e7a}
  \epsilon = \frac{\delta r}{h} = \sqrt \frac{N_P}{N_B} \propto \sqrt{N_P}\frac{\delta n_B}{n_B} .
\end{equation}

\noindent Here $r$ is the space coordinate, $h$ is the cell size, $N_{P}$ is the number of numerical macro-particles which substitute for $N_{B}$ real beam protons in the cell volume, and $n_{B}$ is the local beam density. The random displacements \eqref{e7a} lead to the same level of numerical beam density fluctuations as the $\sqrt{N_B}$  fluctuations of the real beam proton number within the cell volume.

Fig.~\ref{fig5} shows results of the 3D simulation. The frame Fig.~\ref{fig5}(a) gives the wake field growth with propagation distance. The broken line corresponds to the simulation without any external seeding. The proton bunch self-modulation started to grow from the natural beam density fluctuations due to the shot noise. We see that it takes 10~m of plasma for the wake to reach its maximum at about 80~MV/m. At the distance of 10~m, the proton bunch has already significantly diffracted due to its transverse emittance. The generated wake field behind the proton bunch is shown in Fig.~\ref{fig5}(b). We zoom here at the position $2\sigma_z$ behind the middle of the SPS bunch. One sees that the wake is not very symmetric and is tilted. This tilt is due to the presence of the hosing instability that competes with the axisymmetric self-modulational mode \cite{PRL104-255003}. The solid red line in  Fig.~\ref{fig5}(a) shows the wake field growth when the self-modulation is seeded by a precursor electron bunch. We used a 10~MeV electron bunch of 1~ps duration and the current of 100~A as the precursor. It is seen that the field fluctuates strongly at the first meter of propagation in plasma. This is due to strong self-focusing/defocusing of the electron bunch. After the first meter, the self-modulation of the proton bunch sets in and the field starts to grow exponentially. The maximum field of some 0.6 GV/m is reached after 4 meters propagation distance. The field snapshot at this distance is given in  Fig.~\ref{fig5}(c). Apparently, the field is symmetric so that the self-modulation mode wins over the hosing. Our simulations also show that the growth of self-modulation is not very sensitive to the electron bunch parameters such as the energy and the current. It is important that the initial wake generated by the precursor is significantly larger than the noise field of the proton bunch and that the saturation of the self-modulation is reached before the proton bunch diffracts away due to its transverse emittance.

This work is supported by by the Ministry of Education and Science of the Russian Federation, RFBR (grants 11-01-00249, 11-02-00563, and 12-01-00727), grant 11.G34.31.0033 of the Russian Federation Government, and RF President's grant NSh-5118.2012.2.

\end{document}